# Measurement of the infrared complex Faraday angle in semiconductors and insulators

**M.-H. Kim, V. Kurz, G. Acbas, C. T. Ellis, J. Cerne***

*Department of Physics, University at Buffalo,SUNY, Buffalo, New York, 14260, USA*
*jcerne@buffalo.edu

**Abstract:** We measure the infrared (wavelength $\lambda$ = 11 – 0.8 $\mu$m; energy $E$ = 0.1 – 1.5 eV) Faraday rotation and ellipticity in GaAs, $BaF_2$, $LaSrGaO_4$, $LaSrAlO_4$, and ZnSe. Since these materials are commonly used as substrates and windows in infrared magneto-optical measurements, it is important to measure their Faraday signals for background subtraction. These measurement also provide a rigorous test of the accuracy and sensitivity of our unique magneto-polarimetry system. The light sources used in these measurements consist of gas and semiconductor lasers, which cover 0.1 – 1.3 eV, as well as a custom-modified prism monochromator with a Xe lamp, which allows continuous broadband measurements in the 0.28 – 1.5 eV energy range. The sensitivity of this broad-band system is approximately 10 $\mu$rad. Our measurements reveal that the Verdet coefficients of these materials are proportional to $\lambda^{-2}$, which is expected when probing with radiation energies below the band gap. Reproducible ellipticity signals are also seen, which is unexpected since the radiation is well below the absorption edge of these materials, where no magnetic circular dichroism or magnetic linear birefringence should occur. We suggest that the Faraday ellipticity is produced by the static retardance ($R_s$) of the photoelastic modulator (PEM) and other optical elements such as windows, which convert the polarization rotation produced by the sample into ellipticity. This static retardance is experimentally determined by the ratio of the Faraday rotation and ellipticity signals, which are induced by either applying a magnetic field to a sample or mechanically rotating the polarization of light incident on the PEM and/or other optical components.

## References and Links

## 1. Introduction

Faraday measurements probe materials by exploring changes in the polarization of transmitted light. These changes are induced by the presence of a magnetic field or magnetization, and are highly sensitive to critical properties such as electronic band structure and spin population. Faraday measurements have been widely performed on semiconductors [1] and more recently on III-V diluted magnetic semiconductors [2-5] and high temperature superconductors [6,7] in the visible and infrared energy range. These measurements provide detailed new information which complement conventional infrared conductivity measurements [8]. The Faraday effect in many optical materials such as GaAs, $BaF_2$, $LaSrGaO_4$, $LaSrAlO_4$, and ZnSe, is well studied in the visible and near-infrared range. However, similar mid-infrared (MIR, wavelength $\lambda = 11 - 2$ $\mu$m; energy $E = 0.1 - 0.6$ eV) Faraday measurements have been rare. Understanding the nature of these materials in the MIR is extremely important because they are commonly used as substrates, windows, and lenses in studies of more remarkable materials such as, high-temperature superconducting cuprates (HTSC) [6,7] and III-V(Mn) diluted magnetic semiconductors [9-12], which are particularly revealing in this spectral regime. The key features of the measurements presented in this paper are: the first report of the MIR Verdet constant for $BaF_2$ and $LaSrAlO_4$; measurement of both Faraday rotation and ellipticity; and a broad nearly continuous infrared spectral range. Furthermore, since we are developing new instrumentation and techniques, these measurements provide an excellent test of the accuracy and sensitivity of our magneto-polarimetry system.

Magneto-optical transmission measurements with the magnetic field and radiation propagation directions both oriented perpendicular to the sample surface, determine the complex Faraday angle produced by a sample in a magnetic field. The real part of the Faraday angle $\mathrm{Re}(\theta_F)$ is related to the rotation of the plane of polarization (Faraday rotation) and the imaginary part Im($\theta_F$) is connected to the ellipticity (Faraday ellipticity) of the transmitted light [13,14]. In principle, the Faraday effect originates from optical transitions and free carriers in a magnetic field [1]. For an isotropic sample in the Faraday geometry, Faraday rotation results from differing indices of refraction for left- and right-circularly polarized light, where a difference in absorption for left- and right-circularly polarized light produces Faraday ellipticity. In non-isotropic cases, Faraday ellipticity can also arise from stress-induced linear birefringence [15], or linear magnetic birefringence [16].

The MIR energies used in this experiment are up to an order of magnitude smaller than the band gap energy of the semiconductors and insulators studied here. In general, there will be no absorption of light in this energy range. When a magnetic field is applied to a sample the interband transition energies will shift, causing a difference in the indices of refraction for left- and right-circularly polarized light below the band gap. A phase shift in left- and right-circularly polarized waves is induced by this difference which causes a polarization rotation in the transmitted light. This angle of rotation Re($\theta_F$) is proportional to the sample's thickness $d$ and applied magnetic field $H$. The Faraday angle normalized by the field and sample thickness is known as the Verdet coefficient ($V$ = Re($\theta_F$)/($Hd$)). The Verdet coefficient for below band gap radiation in an insulating sample has only interband contributions, and is proportional to $\lambda^{-2}$. As mentioned earlier, Faraday ellipticity is not expected in these materials in the MIR regime, because the radiation energy is below the absorption edge. However, reproducible ellipticity signals proportional to $H$ are found throughout our experiment. Note that Faraday ellipticity is defined as the ratio of the minor to major axes of the polarization ellipse. In this study we suggest that the source of the observed ellipticity arises from the rotation of the polarization incident on optical components after the sample (e.g. PEM, windows, substrates, lenses, etc...). It is important to better understand this ellipticity artifact so that we may isolate the real signal produced by the sample. For example, one similar artifact that we previously found was that stray magnetic fields can induce rotation and ellipticity signals in the output of gas lasers that are comparable to the signals produced by samples [14]. This artifact is easily removed by placing the lasers farther away from the magneto-optical cryostat [14].

In this paper, we present the MIR complex Faraday angle measurements of GaAs, $BaF_2$, $LaSrGaO_4$, $LaSrAlO_4$, and ZnSe. We shall first introduce the magneto-optical measurement setup, focusing on our newly custom-modified prism monochromator. Next, we determine the Verdet coefficients of these samples and analytically derive the static retardance of optical components from the rotation and ellipticity signals. Finally, we discuss new calibration techniques and some important artifacts including the static retardance of our PEMs and other optical components such as windows and polarizers.

## 2. Experimental Technique

There are two ways linearly polarized light can change after interacting with a sample in a magnetic field. 1) The plane of the polarization can rotate and 2) the light can acquire ellipticity. These polarization changes in the transmitted light are characterized by the complex Faraday angle $\theta_F$.

Our gas lasers ($CO_2$; 9 – 11 $\mu$m, CO; 5 – 6 $\mu$m, HeNe; 3.4 $\mu$m) are set up as described in Ref. [14]. However, for this work we have expanded the number of usable laser lines with new semiconductor lasers (wavelengths 2.5, 2.0, 1.5, 1.3, 0.978 $\mu$m). These lasers provide Faraday angle measurements throughout a wide range of discrete energies (0.1 eV – 1 eV). To further expand the capabilities of our system we have also added a new custom-modified double pass prism monochromator (Perkin-Elmer Model 99) with a 300 W Xe lamp (Perkin-Elmer Cermax) which allows continuous broadband measurements in the 0.28 – 1.5 eV energy range. Unlike typical arc lamps, which are housed in glass and therefore cannot be used beyond 2 $\mu$m, our Cermax arc lamp housing contains a sapphire window, which is transparent out to 6 $\mu$m. Furthermore, standard globar sources tend to have large radiating areas compared to the output power and color temperatures near 1000 K (e.g. the 140 W Newport/Oriel model 6363 infrared emitter has a color temperature below 1100 K and a radiating area of 6.4mm × 17.5mm). Our arc lamp has a significantly smaller source area (arc gap is approximately 1 mm and the radiating area is even smaller) and a color temperature of 6000 K. This smaller source size translates into a much brighter illumination spot on the sample (5 mm × 5mm) compared to typical globar sources, which tend to overfill the sample when imaged on to it. Based on Planck's black body radiation law, the six-fold increase in color temperature alone increases the intensity of light from the Cermax arc lamp at wavelengths of 2 $\mu$m and 4 $\mu$m by factors of 600 and 40, respectively. The critical advantage of the double pass prism monochromator is that it relies on a $CaF_2$ prism to disperse the radiation instead of a diffraction grating, which is used in most new monochromators. The prism refracts each wavelength into a unique angle, which is not the case with diffraction gratings. For example, with a grating the first order diffraction peak position for a wavelength of 2 $\mu$m will also contain the 2nd order peak for 1 $\mu$m wavelength light. When scanning wavelengths from 1 – 5 $\mu$m using a grating, one would need to use a series of short-wavelength cut-off filters to remove the higher order peaks from shorter wavelengths (which typically are more intense since the source output intensity drops at longer wavelengths). This is not only cumbersome and expensive, but it also reduces the throughput of the system. It should be noted that for our setup we are reversing the beam path through the monochromator in order to use it as a source instead of an analyzer. i.e., light exits the original entrance slit (Slit 2 in Fig. 1).

Figure 1 shows the optical path of our broadband light source (a detailed description of the gas laser beam path is found in Ref. [14]). Firstly, light from the Xe arc lamp is focused on to Slit 1. The light is then collimated by a parabolic mirror and sent to the $CaF_2$ prism. The light passes through the prism four times before exiting the monochromator at Slit 2. The desired probe wavelength is chosen by rotating the prism. The wavelength of light that exits Slit 2 has been calibrated in the 4.4 $\mu$m – 0.95 $\mu$m range by a Bomen Fourier transform infrared spectrometer using an InSb detector. A grating monochromator and photomultiplier tube detector were used to calibrate wavelengths from 0.95 $\mu$m – 0.63 $\mu$m (the grating monochromator was calibrated using Hg and Kr lamps). The output beam of the broadband source has a Gaussian profile with respect to wavelength as shown in the inset of Fig. 1. The spectral linewidth in the 4.4 $\mu$m – 0.95 $\mu$m range is 50 nm or less using 0.05 inch slit widths. For the 0.95 $\mu$m – 0.63 $\mu$m region the linewidth is about 20 nm with 0.02 inch slit widths. After emerging from the monochromator the light is vertically polarized (i.e. oriented along *x*-axis) by polarizer $P_1$ as shown in Fig. 1.

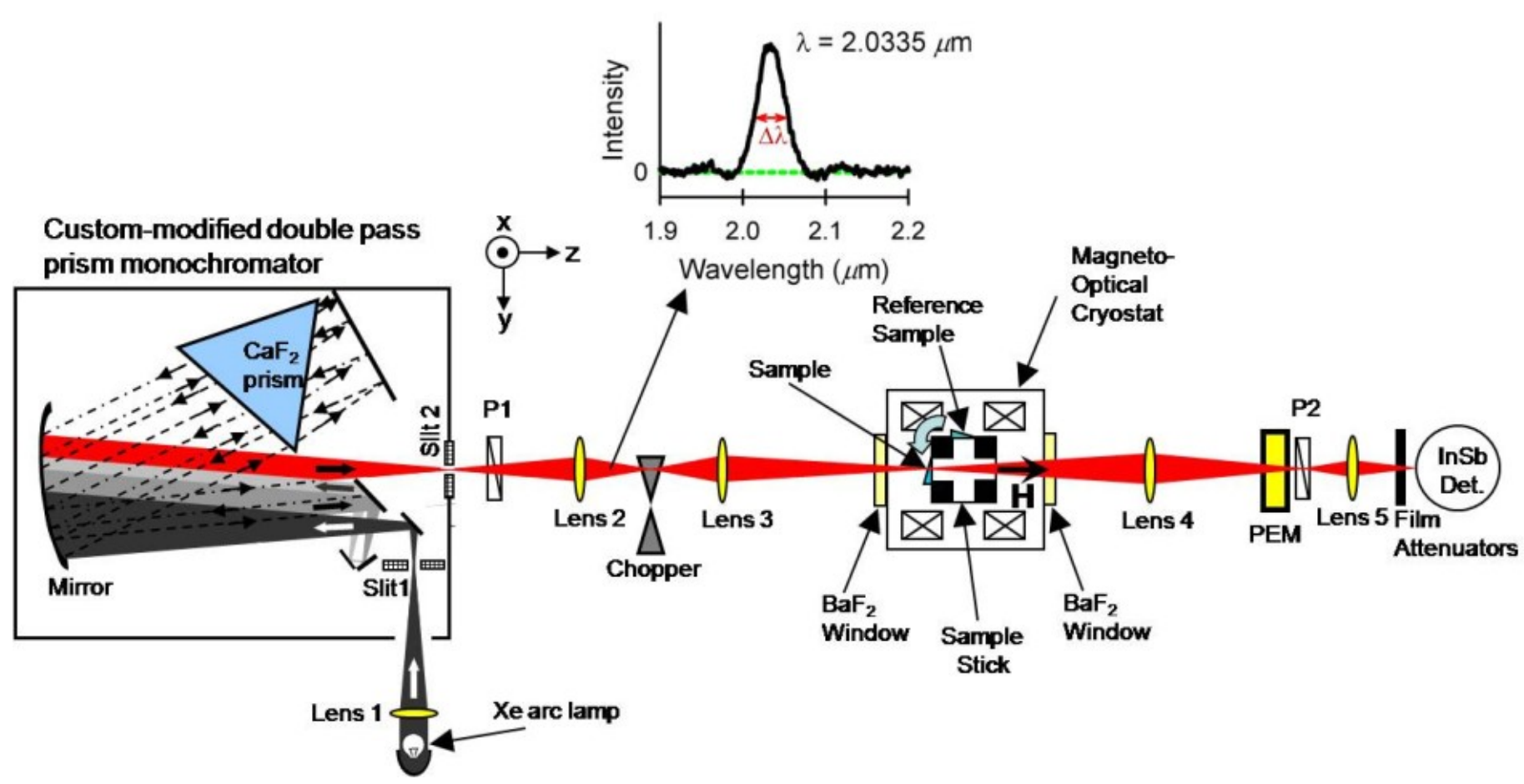

Fig. 1 Overall schematic of the optical path, as viewed from above. The beam passes the $CaF_2$ prism four times: dashed line with arrows (1st and 2nd time) and dot-dashed line with arrows (3rd and 4th time). The wavelength exiting the monochromator is selected by rotating the prism. The exit beam has a Gaussian intensity profile as a function of wavelength, as shown for 2 $\mu$m in the inset. $\Delta\lambda$ represents the spectral linewidth. The sample stick is rotatable to select sample or reference sample.

The sample is located in a magneto-optical cryostat which can reach temperatures down to 6 K and magnetic fields up to 7 T. When using gas lasers, the cryostat is fitted with two ZnSe windows mounted on 30 cm extension tubes. This large distance from the cryostat reduces Faraday rotation produced in the windows by stray magnetic fields interacting with the windows. The sample substrate and the ZnSe windows are wedged 1° – 2° to prevent étalon artifacts from multiple reflections. For the broadband setup unwedged 2.5 inch $BaF_2$ windows are attached directly to the cryostat tail piece without any extension tubes. We expect that incoherent broadband light does not suffer much from multiple reflections in the $BaF_2$ windows. The stray magnetic field is estimated to be approximately 0.01 T at the ZnSe windows and 0.3 T at the $BaF_2$ windows when the field at the center of the magnet is 1 T. We are currently fabricating extension tubes for the $BaF_2$ windows to reduce the background rotation produced by these windows.

The photoelastic modulator (PEM) modulates the phase of the two orthogonal linear polarization components that pass through it. It is this modulation that allows us to determine the polarization of the beam that passes through the sample. The types of PEMs and polarizers used in the system depend on the probe wavelength. In the 4.44–2 $\mu$m wavelength range, we use a ZnSe PEM (II/ZS50, Hinds Instruments) in combination with two $BaF_2$ holographic wire-grid polarizers (Thorlabs WP2SH-B) labeled as $P_1$ and $P_2$, respectively in Fig. 1. For the 2–0.63 $\mu$m wavelength range we use a fused silica PEM (I/FS50, Hinds Instruments) and calcite Glan-Taylor polarizers (MGTYA 20, Karl Lambrecht Corporation). To keep multiple reflections within the PEM from reaching the detector the ZnSe PEM is tilted forward 25° and the fused silica PEM crystal is wedged. The optical axis of the PEM is oriented along the *x* axis, which is the same orientation as the incident light polarization shown in Fig. 1. After passing through the sample, the transmitted light acquires a small *y*-component of polarization, which produces rotation and ellipticity. The PEM modulates the phase of this *y*-component of polarization with respect to the *x*-component at a frequency $\omega_{\mathrm{PEM}} \approx 50 \times (2\pi)$ kHz. The PEM modulates the phase difference between the *x* and *y* components of the transmitted light sinusoidally: $\delta(t) = R_d \cos(\omega_{\mathrm{PEM}} t)$, where $R_d$ is the dynamic retardance and is the phase modulation amplitude of the PEM. A linear polarizer ($P_2$) placed after the PEM is oriented at $\alpha_2$=45° with respect to the *x* axis to mix the *x* and *y* polarization components of the light exiting the PEM.

The intensity of the modulated light is measured by a liquid-nitrogen-cooled detector. Mercury-cadmium-telluride (MCT) and InSb detectors are used for the lasers and the broadband light source, respectively. Three lock-in amplifiers are used to obtain the complex Faraday angle. One lock-in amplifier is referenced to the chopper frequency $\omega_0$ (~ 1 kHz) in order to measure the overall light intensity. The two others are locked onto the even and odd harmonics of $\omega_{\mathrm{PEM}}$ in order to detect the polarization of the beam. The even harmonics of $\omega_{\mathrm{PEM}}$ are related to the rotation $\mathrm{Re}(\theta_F)$ while the odd harmonics are related to the ellipticity $\mathrm{Im}(\theta_F)$ [13]:

$$\mathrm{Re}(\theta_F) = \frac{I_{2\omega_{\mathrm{PEM}}}}{4J_2(R_d)I_0}, \quad \mathrm{Im}(\theta_F) = \frac{I_{3\omega_{\mathrm{PEM}}}}{4J_3(R_d)I_0}, \tag{1}$$

where $J_n$ is the *n*th order Bessel function, $I_{n\omega_{\mathrm{PEM}}}$ is the intensity of the light modulated at $n\omega_{\mathrm{PEM}}$, and $I_0$ is the intensity of light modulated at the chopper frequency.

All samples are polished with a wedge in order to separate internal multiple reflections. The amount of wedging depends on the material. The wedge angle and thickness of our samples are as follows: for the 0.51 mm thick GaAs sample, a 1° wedge is used, while the $BaF_2$, $LaSrGaO_4$, $LaSrAlO_4$ samples are wedged by 2° with average thicknesses of 0.81 mm, 0.38 mm, and 0.40 mm, respectively. Faraday measurements are also performed on the windows without a sample. Two ZnSe windows are wedged 2°, each with an average thickness 5 mm. All of these samples/windows show Faraday rotation as well as ellipticity in the MIR energy range.

The Verdet coefficient has two terms. The first term is due to the interband absorption resonance and scales with the light wavelength as $1/\lambda^2$. The second term is due to free carrier interaction with radiation and scales as $\lambda^2$. For below band gap radiation the Verdet coefficient can be expressed as [17]

$$V(\lambda) = \frac{\mathrm{Re}[\theta_F(\lambda)]}{Hd} = \frac{u}{\lambda^2} + v\lambda^2, \tag{2}$$

where $u$ is the coefficient related to the interband contribution, and $v$ is the free carrier contribution coefficient. The insulating samples probed in this experiment have extremely small free carrier densities, thus the interband transition term dominates, and one can expect the Verdet coefficients of the different semiconductors and insulators to have the same slope in a log-log plot of $V(\lambda)$.

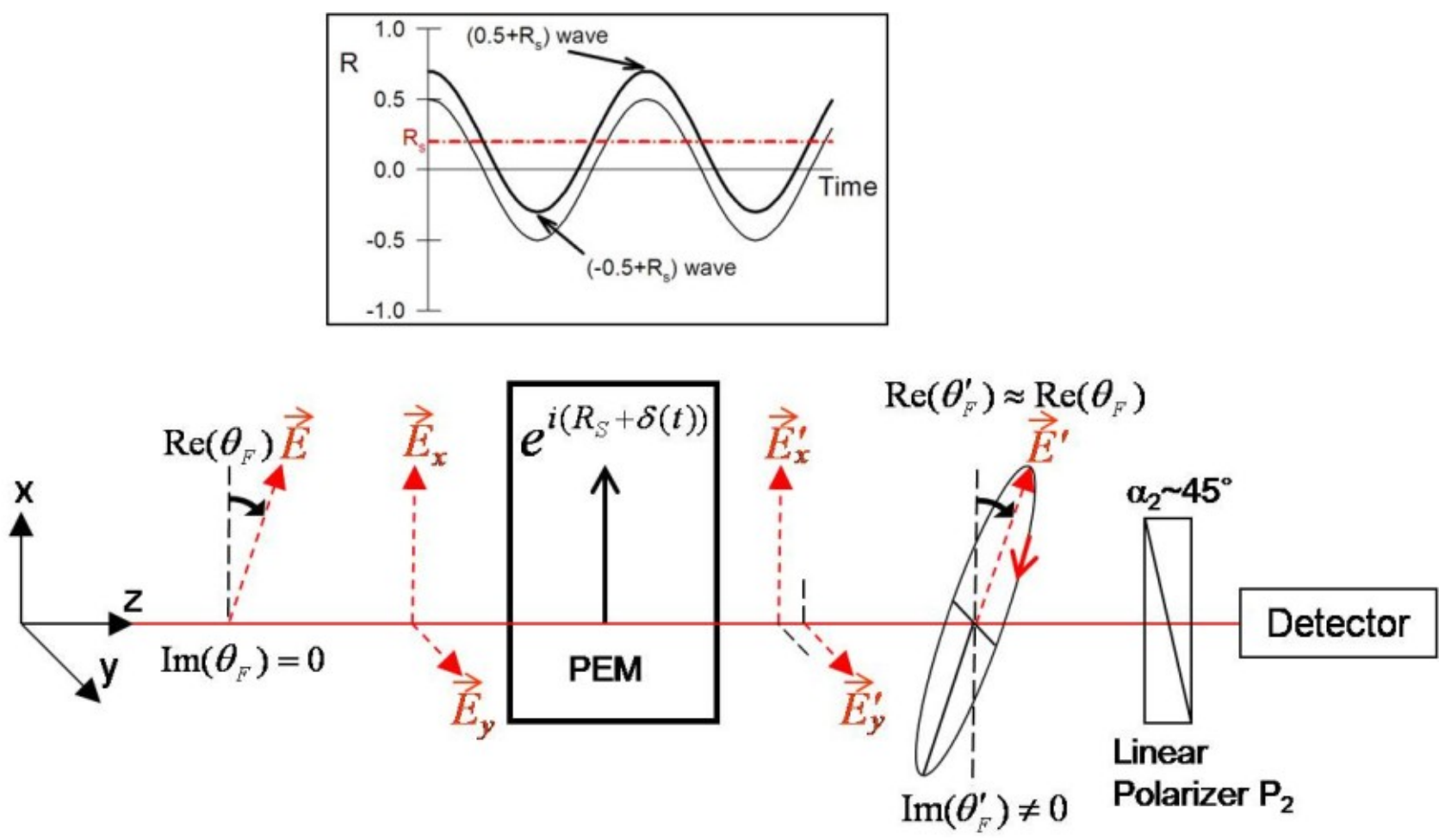

Fig. 2. Optical path of the light after passing through the sample. Dashed arrows show the electric field vector. The time dependent PEM dynamic retardance (thin line) is shown in the boxed inset when one sets the PEM retardance to 0.5 wave. Dot-dashed line in the inset represents the PEM's static retardance $R_s$. The total retardance (thick line) is shifted by $R_s$.

For below band gap radiation, the MIR Faraday ellipticity of these materials will be negligibly small ($\mathrm{Im}(\theta_F) \approx 0$) because there is no absorption. However, Faraday ellipticity is commonly observed due to other components in the system (e.g. as previously mentioned stray magnetic fields interacting with gas laser can produce Faraday ellipticity [14]). Ellipticity can also be produced by the PEM's static stress-induced birefringence. In this case the PEM acts like a static wave plate with a static retardance $R_s$ caused by non-homogeneous mounting stresses on the PEM crystal. This static retardance causes a shift in the dynamic retardance $R_d$ by a constant as shown by the inset in Fig. 2. Note that the optical axis of the static retardance may be oriented differently than the dynamic retardance axis, but in this paper we simply consider the case where the two optical axes have the same orientation. The net retardance of the PEM is given by:

$$R_{\mathrm{PEM}}(t) = \delta(t) + R_s \tag{3}$$

We assume that the light transmitted by the sample experiences Faraday rotation only (i.e., $\mathrm{Im}(\theta_F) = 0$). After passing through the sample, the PEM, and $P_1$, the light containing signals $I_0$ and $I_{\omega_{\mathrm{PEM}}}$ is incident on the detector. The lock-in amplifiers then pick off the intensity of the two PEM harmonics with frequencies $2\omega_{\mathrm{PEM}}$ and $3\omega_{\mathrm{PEM}}$. The intensity ratio of the two signals produces

$$\frac{I_{3\omega_{\text{PEM}}}}{I_{2\omega_{\text{PEM}}}} = \frac{J_3(R_d)}{J_2(R_d)}\tan(R_s). \tag{4}$$

Equation (4) is equivalent to the static retardance formula obtained from the intensity ratio of $I_{\omega_{\text{PEM}}}$ and $I_{2\omega_{\text{PEM}}}$ in Ref. [15]. The ratio $\text{Im}(\theta_F)/\text{Re}(\theta_F) = (I_{3\omega_{\text{PEM}}} J_2(R_d))/(I_{2\omega_{\text{PEM}}} J_3(R_d))$ shown in Ref. [13], allows Eq. (4) to be simplified as

$$\text{Im}(\theta_F) = \text{Re}(\theta_F)\tan(R_s). \tag{5}$$

Equation 5 shows that the real part of the Faraday angle is connected to the imaginary part by the PEM's static retardance. This implies that any optical element with a static retardance, such as a PEM, placed after the sample can convert a pure rotation in polarization into ellipticity. Furthermore, the static retardance can be expressed as $\Delta L/\lambda$, where $\Delta L$ is the optical path difference produced by the mounting stress. Although the effective $\Delta L$ can depend on wavelength [18], we assume that it is constant here to obtain the first order behavior of $\text{Im}[\theta_F(\lambda)]$. For below bandgap radiation in insulating samples, $\text{Im}(\theta_F)$ becomes:

$$\text{Im}[\theta_F(\lambda)] = \frac{u}{\lambda^2}\tan\left(2\pi\frac{\Delta L}{\lambda}\right) \approx \frac{\Delta L}{\lambda^3}, \tag{6}$$

where we have assumed that $\Delta L/\lambda << 1$. Therefore, one sees that the Faraday ellipticity artifact originates from Faraday rotation, and that this ellipticity will roughly scale as $1/\lambda^3$.

### 3. Verdet coefficients

We measure Verdet coefficients for materials that are commonly used as substrates and windows over a wide energy range from MIR to visible (0.1 – 1.5 eV). Ellipticity artifacts caused by the static retardance of the PEM and other optical components are also described. The samples do not induce ellipticity in the MIR. However, our measurements and analysis confirm that the PEM's static retardance can translate rotations into ellipticity signals.

Figure 3 shows $\theta_F$ for GaAs, $BaF_2$, $LaSrGaO_4$, and ZnSe normalized by the magnetic field $H$ as a function of wavelength $\lambda$ in a log-log plot. These Faraday angles are measured using both laser and the broadband sources for GaAs and $BaF_2$, whereas only lasers are used for the other samples. $\theta_F$ produced by these samples increases as the wavelength shortens. For $\text{Re}(\theta_F)$, there is a difference of approximately two orders of magnitude between longer and shorter wavelengths: $10^{-4}$ to $10^{-2}$ rad for GaAs, and $10^{-5}$ to $10^{-3}$ rad for the other materials. The $\text{Im}(\theta_F)$ varies by three orders of magnitude for GaAs ranging from $10^{-5}$ to $10^{-2}$ rad. The noise floor for both $\text{Re}(\theta_F)$ and $\text{Im}(\theta_F)$ is approximately $10^{-5}$ rad, so the decay of these signals at longer wavelengths is lost in the noise for the $BaF_2$, $LaSrGaO_4$, and ZnSe samples. For comparison, typical Faraday signals for non-ferromagnetic metals are on the order of $10^{-4}$ rad at 8 T and 100 meV [6,19]. For ferromagnetic metals such as the ruthenate perovskite $SrRuO_3$ and the diluted magnetic semiconductor $Ga_{1-x}Mn_xAs$ films, the Faraday angles are on the order of $10^{-2}$ rad at 1 T from 0.1 – 0.8 eV [14]. At longer wavelengths (~ 10 $\mu$m), Faraday angles from ferromagnetic metallic films (~ 100 nm thick) usually dominate the signal, but the $\text{Re}(\theta_F)$ from the substrates and windows becomes more important as the wavelength is decreased. The vertical dashed line at $\lambda$ = 2 $\mu$m in Fig. 3 marks the boundary of the two different PEM/polarizer sets. Interestingly, the $\text{Re}(\theta_F)$ is continuous across the boundary, while the $\text{Im}(\theta_F)$ shows a clear discontinuity at the boundary. However, the slope for $\text{Im}(\theta_F)$ is similar on both sides of the boundary. This implies that the ellipticity does not come from the sample itself, but rather it is an artifact caused by the PEM, polarizers, and/or windows, which are different in the two measurement ranges.

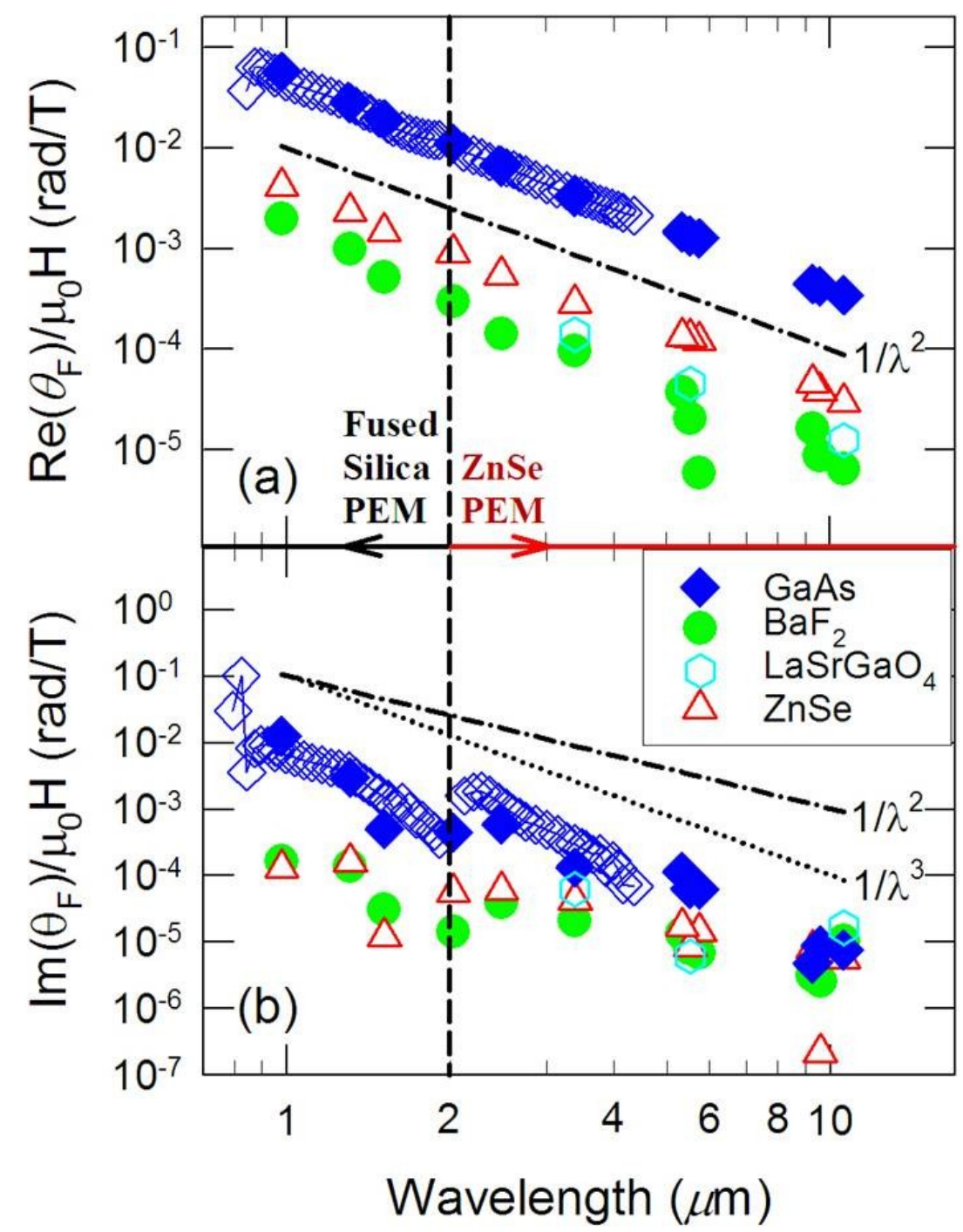

Fig. 3. The log-log plot of the (a) Re($\theta_F$) and (b) Im($\theta_F$) divided by the magnetic field $H$ as a function of wavelength $\lambda$. The vertical dashed line divides the regions where the two different PEM/polarizers are used. For GaAs, the solid diamonds represent laser measurements and open diamonds represent broadband measurements. Dot-dashed line and dotted line are guides to the eye indicating $1/\lambda^2$ and $1/\lambda^3$ dependence, respectively.

The wavelength dependence of the Verdet coefficient Re($\theta_F$)/$H$ is shown in Fig. 3(a). According to Eq. (2) the slope of these data determine to which power $\lambda$ is raised in Eq. (2), which is expected to be -2 for below band gap radiation in insulating materials. Table 1 shows the gap energies of the samples, which are larger than the probe energies used in this experiment (0.1 – 1.5 eV). The only exception is for GaAs, which has an energy gap of 1.43 eV. This results in the data deviating from Eq. (2) near the band gap energy of GaAs in Figs. 3 and 4. The dot-dashed reference line in Fig. 3 acts as a guide indicating a $1/\lambda^2$ dependence. All samples produce a slope near -2 as shown in Table 1, but the slope of the $BaF_2$ data deviates more than the others, probably due to the noise from the weak Faraday signals ($10^{-5}$ rad, at 1 T) at wavelengths longer than 5 $\mu$m. Faraday measurements using the broadband light source in GaAs (open diamonds) cover the regions between laser points (solid diamonds) which demonstrates that the various light sources are consistent with one another.

**Table 1. Verdet coefficients ($V(\lambda) \propto \lambda^{\alpha}, V(\lambda) = u/\lambda^2$)**

| Material | $E_{gap}$ (eV) | α (Theory α= -2) | $u$ (°$\mu$m$^2$/Tm) | $u_{10.5\mu m}$ (°/Tm) | Ref. $u_{10.5\mu m}$ (°/Tm) |
|---|---|---|---|---|---|
| GaAs | 1.43 [20] | -2.108 | 5725.3 | 51.93 ± 9.56 | 44 [13] |
| $BaF_2$ | 11.0 [21] | -2.523 | 110.7 | 1.00 ± 0.42 | |
| $LaSrGaO_4$ | 2.85 – 2.87 [22] | -2.086 | 195.4 | 1.77 ± 0.13 | 1.5 [13] |
| $LaSrAlO_4$ | 2.84 – 3.0 [22] | -1.873 | 72.1 | 0.65 ± 0.34 | |
| ZnSe | 2.7 [23] | -2.010 | 2241.6 | 20.3 ± 6.1 | 24.4 [24] |

To further clarify the wavelength dependence of the Verdet coefficient, one can plot $V(\lambda)$ as a function of $1/\lambda^2$. Figure 4 shows $V(\lambda) = \mathrm{Re}(\theta_F)/(Hd)$ from GaAs as a function of $1/\lambda^2$. As expected the data shows a linear dependence for below gap radiation and becomes non-linear when the gap energy is approached.

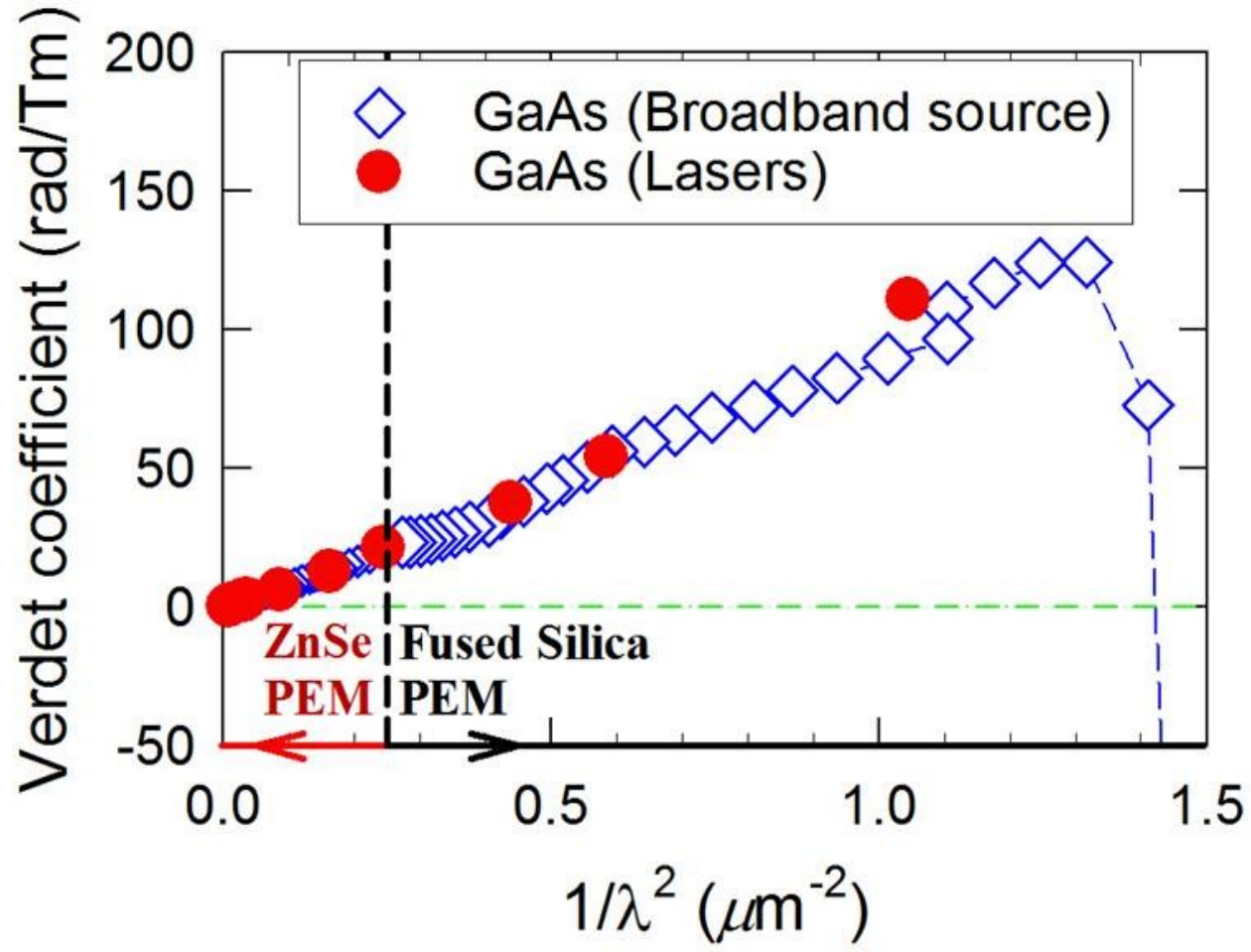

Fig. 4. Verdet coefficient of GaAs as a function of $1/\lambda^2$. Note that the wavelength dependence deviates for $1/\lambda^2$ near the bandedge.

The constant $u$ in Eq. (2) is related to the energy gap and can be determined either by the intercept in Fig. 3(a) or by the slope in Fig. 4. For Table 1, we use the latter method, but these two methods agree within 10%. The units of $u$ are $(°\mu m^2)/(Tm)$, however one usually considers the Verdet coefficient at a specific wavelength, so the constant $u$ often has units °/(Tm) at a particular wavelength. Typically, lower gap energies produce larger values of $u$. For example, GaAs, with an energy gap half as large as that of ZnSe, has double the Verdet constant. Likewise, $BaF_2$, which has an energy gap that is an order of magnitude larger than the gaps in ZnSe and GaAs, has a Verdet coefficient that is an order of magnitude smaller. However, $LaSrGaO_4$, $LaSrAlO_4$, and ZnSe have almost identical gap energies, but $u$ for ZnSe is an order of magnitude larger. Our measurements of $u$ are in good agreement with other published data at 10.5 $\mu$m as can be seen in Table 1. The ZnSe measurements were only made with windows rather than samples made of this material. The windows are located a significant distance away from the center of the magnet (~ 60 cm), therefore the magnetic field had to be estimated using the stray field plots for our magnet. This leaves us with a fairly large uncertainty (±30%) in the Verdet values for ZnSe. To the authors' best knowledge, this work is the first reported measurement of the MIR Verdet coefficients for $BaF_2$ and $LaSrAlO_4$. Note that 12°/Tm reported for $LaSrGaO_4$ in Ref. [13] is a typographical error, which should read 12°/m at 8 T.

Figure 3(b) shows $\mathrm{Im}(\theta_F)$, which like $\mathrm{Re}(\theta_F)$ scales as a power of $\lambda$. In Eq. (6), $\mathrm{Im}(\theta_F)$ has a $1/\lambda^3$ dependence when the PEM acts like a static waveplate, thereby producing an ellipticity. When compared to the reference lines in the plot it is quite clear that the GaAs data has a $1/\lambda^3$ dependence (except near 2 $\mu$m as will be discussed later). There is also a clear discontinuity in these data which result from the different PEM's having different static retardances. The slopes, however, are similar in the two regions. Unfortunately, ellipticity from the other samples is noisy due to weak Faraday signal near $10^{-5}$ rad, but the data are consistent with $1/\lambda^3$ below 2 $\mu$m where the magnitude of the signal is well above $10^{-5}$ rad. The ellipticity artifact is approximately two orders of magnitude smaller than the rotation signal, but can be important in materials which produce large Faraday rotation and small Faraday ellipticity.

## 4. Discussion

In this section, we introduce a retardance calibration technique, which provides a more reliable polarimetry system calibration. We also discuss the sensitivity of Faraday measurements using our system, and explore ellipticity artifacts.

### *4.1 Calibration*

We previously developed a calibration technique to simultaneously determine both $R_d$ and the orientation angle $\alpha_2$ of the final linear polarizer $P_2$ [14]. The calibration and all measurements are performed near $R_d = 2.406$ rad with $\alpha_2 = 45°$. The calibration is performed by rotating the PEM and $P_2$ by a known angle as a single unit. The changes in the $2\omega_{PEM}$ and $4\omega_{PEM}$ signals allows us to determine $R_d$ and $\alpha_2$.

In this study we find that this calibration technique produces errors (as indicated by variations in $\alpha_2$, which should remain constant) when the extinction ratio of the linear polarizers in the system is less than 100:1. Therefore, we developed an independent test to determine $R_d$. We mount a static wave plate with retardance $R_W$ on a rotating stage in front of the PEM. The stage is then rotated with frequency $\omega_0 \approx 27 \times (2\pi)$ Hz, which is much lower than the PEM modulation frequency $\omega_{PEM} = 50 \times (2\pi)$ kHz. Equations (14a) and (14b) in Ref. [13] reveal that the intensity $I_{4\omega_0}$ at the 4th harmonic of $\omega_0$ can be expressed as a function of $R_d$,

$$I_{4\omega_0}(R_d) = \frac{1}{2} J_0(R_d)(1 - \cos(R_W)), \tag{7}$$

where $J_0$ is the zeroth order Bessel function. One can adjust $R_d$ of the PEM while spinning the wave plate to find the PEM retardance where $I_{4\omega_0}(R_d) = 0$, which is also where $J_0(R_d) = 0$. Since $I_{4\omega_0}(R_d) = 0$ when $J_0(R_d) = 0$, regardless of the value of $R_W$, the key advantage of this technique is that its accuracy does not depend on how well $R_W$ is known. The only requirement is that the $R_W \neq 0$. In our case we used a waveplate with $R_W = 0.25$ wave at 0.8 $\mu$m.

These two calibration techniques provide a more reliable determination of $R_d$ for our polarimetry system. When using light with a wavelength of 1.3 $\mu$m or shorter we find that the actual retardance of the fused silica PEM is accurate to within 5% when set to 2.406 rad. It was found that when using wavelengths longer than 1.3 $\mu$m the PEM reaches a physical upper limit before getting to the desired 2.406 rad. The maximum optical path difference for the fused silica PEM is $\Delta L_{max} = 0.49$ $\mu$m. The retardance which saturates the PEM can be determined by $R_d^{max}(\lambda) = (2\pi\Delta L_{max})/\lambda$. Any set retardance that is greater than $R_d^{max}$ will default to this maximum value.

Conversely, the ZnSe PEM cannot drive any retardance lower than $R_d^{min}$ (i.e., the retardance hits a floor and the PEM crystal cannot be driven at smaller amplitudes). The minimum optical path difference for the ZnSe PEM is $\Delta L_{min} = 0.401$ $\mu$m. This corresponds to a critical wavelength of 1.047 $\mu$m where the retarance of 2.406 rad is still attainable. The ZnSe PEM had a retardance accuracy that is comparable to the fused silica. For wavelengths greater than 1.047 $\mu$m, the set 2.406 rad retardance is accurate to within 5%. The ZnSe PEM also has a upper limit to the retardance, however, in this paper we are not using wavelengths where this upper limit would be reached.

Since the angle between the PEM and $P_2$ is kept constant throughout all measurements, the calibrated angle $\alpha_2$ should remain constant at 45° for all laser wavelengths. For the broadband light source, one has even stronger expectations that the calibrated $\alpha_2$ must be the same for all wavelengths because the optical alignment does not change while varying the wavelength, which is not strictly true when different lasers are used. Below 2 $\mu$m with the fused silica PEM, the calibrated angle $\alpha_2$ is 45° ± 3% for all wavelengths. For the ZnSe PEM, $\alpha_2$ is found to be 45° ± 5% for all wavelengths except between 1.36 $\mu$m to 2.21 $\mu$m. In order to maintain a high extinction ratio, wire-grid polarizers (at the spectrometer exit slit and after the PEM) are used for wavelengths greater than 1.78 $\mu$m and calcite Glan-Taylor polarizers are used below 2 $\mu$m. Part of the problem in using the ZnSe PEM below 2 $\mu$m is that its anti-reflection coating, which is designed for 9 – 11 $\mu$m wavelength radiation, strongly absorbs radiation around 2 $\mu$m. The poor calibration accuracy in certain wavelength ranges is strongly correlated to poor polarizer extinction ratios and low light intensity, such as near 2 $\mu$m with the ZnSe PEM. We have also seen irregular behavior in $\alpha_2$ with the calcite polarizers as the extinction ratio decreases. The extinction ratio for the calcite polarizer drops below 60:1 in the 1.36 $\mu$m to 2.21 $\mu$m wavelength range. In this wavelength range, $\alpha_2$ obtained from calibration measurements, linearly increases from 45° to 55° as the extinction ratio decreases, despite the fact that $P_2$ is held at a fixed angle. Near 3 $\mu$m, where the calcite polarizer has a good extinction ratio again, $\alpha_2$ is found to be at its nominal value of 45°.

It is important to also realize that calcite polarizers have a limited acceptance angle for which incident light is properly polarized. This acceptance angle is asymmetric and depends on the wavelength. At shorter wavelengths around 0.66 $\mu$m, our calcite polarizers have symmetric acceptance angles of approximately 4° away from normal incidence. However, at longer wavelengths, the acceptance angles are asymmetric. For 2 $\mu$m, one of the acceptance angles is 2° and the other is 6° from normal. The wavelength dependent acceptance angles could allow parts of the beam that were inside the acceptance cone at one wavelength to be outside the acceptance cone at a different wavelength thereby creating a poor extinction ratio and a poor calibration.

*4.2 Sensitivity of the polarimetry system*

Figure 3 shows that data scatters below $10^{-5}$ rad due to noise. Noise can be estimated by rotating the PEM by a known angle. For lasers, the noise is less than $5 \times 10^{-5}$ rad for all wavelengths. Since the broadband light source is much less intense than the lasers, the sensitivity is closer to $10^{-4}$ rad.

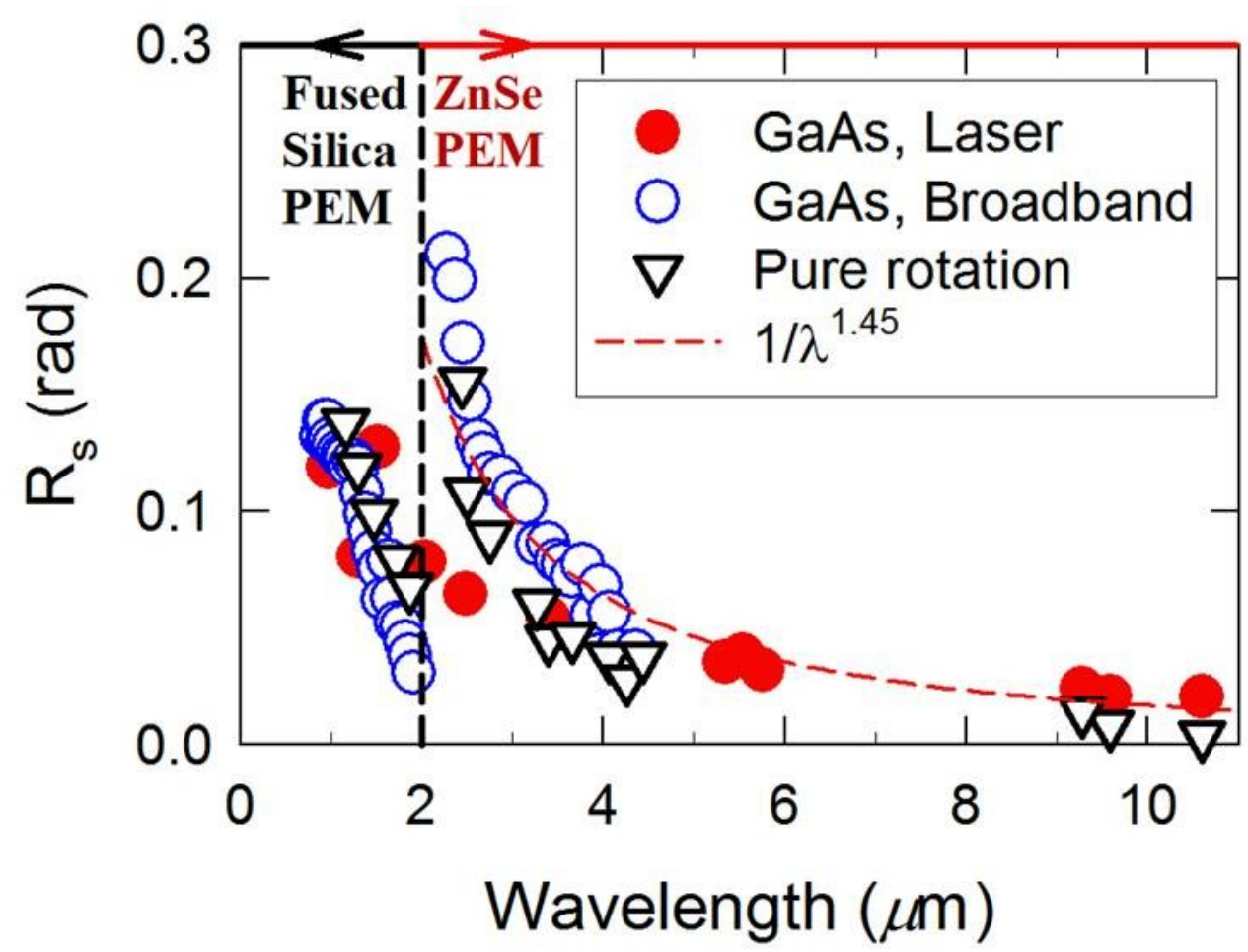

Fig. 5. The static retardance $R_s$ determined from Faraday rotation produced by the GaAs sample in a magnetic field (circles) and from the mechanical rotation of the PEM/polarization without a magnetic field (triangles) as a function of wavelength. The red dashed line is a fit performed above 2 $\mu$m.

*4.3 Artifacts*

As mentioned earlier the most striking artifact that we discovered is the ellipticity signal that arises from real polarization rotation. This artifact is manifested in a Im($\theta_F$) that is proportional to $1/\lambda^3$ and is caused by optical elements in the system behaving like a static waveplate with retardance $R_s$. Equation (5) shows the connection between the real and imaginary parts of the Faraday angle through $R_s$. We have determined this static retardance $R_s$ experimentally by taking the ratio of $\mathrm{Im}(\theta_F)$ and $\mathrm{Re}(\theta_F)$ when the PEM or the laser polarization axis is rotated. Figure 5 shows the static retardance $R_s$ as a function of wavelength. In GaAs, the values of $R_s$ determined by the laser and broadband light sources are consistent with each other. In Eq. (6), we see that $R_s$ follows a $1/\lambda$ dependence. However, the best fit for our data (represented by the dashed line in Fig. 5) reveals a $1/\lambda^{1.45}$ behavior for the ZnSe PEM. It is possible that our measured value of the exponent differs from the expected value of 1 because $\Delta L$ also has a wavelength dependence. The static retardances of the two different PEMs are different from each other since they are composed of different materials that are mounted in different housings. In order to ensure that the Faraday ellipticity is not produced by the sample, one can perform the same tests by making pure mechanical rotations of the PEM and $P_2$ by a small angle $\varphi$ as one unit without a sample and no magnetic field. This is equivalent to the laser polarization axis rotating the same amount in the opposite direction ($-\varphi$). In Fig. 5, the open triangles represent the static retardance deduced by this pure mechanical rotation of the PEM and $P_2$. The static retardance of the ZnSe and fused silica PEMs that are determined by pure rotations are consistent with the ellipticity induced by Faraday rotation from the sample in a magnetic field. We suspect that the anomalous behavior of the measured static retardance for the ZnSe PEM below 2.21 $\mu$m is related to the strong absorption of radiation by its anti-reflection coating in this range.

## 5. Conclusion

We have presented complex Faraday angle measurements in GaAs, $BaF_2$, $LaSrGaO_4$, $LaSrAlO_4$ and ZnSe in the MIR energy range well below the interband absorption edge. The wavelength dependence of the Faraday rotation agrees well with theory for a Verdet coefficient that is dominated by interband transitions. The constant $u$ for the interband contribution in each sample is consistent with the values reported by others at 10.5 $\mu$m. We suggest that the Faraday ellipticity in this experiment is an artifact resulting from Faraday rotation and anisotropic strain in the optical components after the sample. These components act as weak static waveplates, causing ellipticity changes when the sample produces Faraday rotation. The Faraday ellipticity seen in GaAs follows a $1/\lambda^3$ dependence, which is predicted by a simple calculation. The static retardance is obtained from the normalized signal ratio between the Faraday rotation and ellipticity. Furthermore, mechanical polarization rotations produce ellipticity signals comparable to those produced by rotations due to an applied magnetic field. These results are critical for removing the background contributions from GaAs, $BaF_2$, $LaSrGaO_4$, $LaSrAlO_4$, and ZnSe substrates and windows in MIR Faraday measurements. They also confirm the high accuracy of these measurements over the entire MIR wavelength range and indicate that pure rotation signals can produce false ellipticity signals.

## Acknowledgments

We wish to thank H.D. Drew and R.L. Greene for providing us with $LaSrGaO_4$ and $LaSrAlO_4$ samples, B.D. McCombe and A.V. Stier for their assistance in performing the wavelength calibration of broadband light source using their Fourier transform spectrometer, A. Petrou for supplying Hg and Kr lamps for wavelength calibration, and B. Wang from Hinds instruments for helpful discussions about PEMs. This work was supported by Research Corporation Cottrell Scholar Award, NSF-CAREER-DMR0449899, and an instrumentation award from the University at Buffalo, College of Arts and Sciences.